%% file: main.tex
\documentclass[10pt,conference]{IEEEtran}
\IEEEoverridecommandlockouts
\usepackage{cite}
\usepackage{amsmath,amssymb,amsfonts}
\usepackage{algorithmic}
\usepackage{graphicx}
\usepackage{textcomp}
\usepackage{xcolor}
\usepackage{multirow}
\usepackage{booktabs}
\usepackage{enumitem}
\usepackage{soul}
\usepackage{ulem}
\usepackage{diagbox} 
\usepackage[most]{tcolorbox}  
\usepackage{color}
\usepackage{xcolor}
\usepackage{tikz}
\usepackage{tikz-qtree}
\usepackage{pgfplots}
\usepackage{varwidth}
\usepackage{flushend}
\usetikzlibrary{patterns}
\usepackage{url}
\usepackage{subfigure}
\usepackage{threeparttable}
\def\BibTeX{{\rm B\kern-.05em{\sc i\kern-.025em b}\kern-.08em
    T\kern-.1667em\lower.7ex\hbox{E}\kern-.125emX}}

\begin{document}

\title{APIGen: Generative API Method Recommendation}
\author{
\IEEEauthorblockN{Yujia Chen\IEEEauthorrefmark{2},
Cuiyun Gao*\IEEEauthorrefmark{2}\thanks{*Corresponding author. The author is also affiliated with Peng Cheng Laboratory and Guangdong Provincial Key Laboratory of Novel Security Intelligence Technologies.},
Muyijie Zhu\IEEEauthorrefmark{2},
Qing Liao\IEEEauthorrefmark{2},
Yong Wang\IEEEauthorrefmark{4}, 
Guoai Xu\IEEEauthorrefmark{2}
} \\
\IEEEauthorblockA{
\IEEEauthorrefmark{2}Harbin Institute of Technology, Shenzhen, China \\
\IEEEauthorrefmark{4}Anhui Polytechnic University, China
} \\
\IEEEauthorblockA{
\{yujiachen, zhumuyj\}@stu.hit.edu.cn, \{gaocuiyun, liaoqing, xga\}@hit.edu.cn, yongwang@ahpu.edu.cn}
}


\newcommand{\tool}{APIGen}
\newcommand\etal{{\it{et al.\ }}}
\newcommand{\yun}[1]{\textcolor{magenta}{#1}}
\newcommand{\yingli}[1]{\textcolor{blue}{#1}}
\newcommand{\yujia}[1]{\textcolor{blue}{#1}}
\maketitle

\begin{abstract}
    \input{sections/0_abs}
\end{abstract}


\section{Introduction}
    \input{sections/1_intro}
    
\section{Background}  \label{sec:background}
    \input{sections/2_background}

\section{Approach} \label{sec:approach}
    \input{sections/3_approach} 
 
\section{Experimental Setup} \label{sec:setup}
    \input{sections/4_setup}

\section{Evaluation} \label{sec:results}
    \input{sections/5_result}

\section{Discussion} \label{sec:discussion}
    \input{sections/6_discussion} 

\section{Related Work}  \label{sec:related}
    \input{sections/7_relatedwork}

\section{Conclusion} \label{sec:conclusion}
    \input{sections/8_conclusion} 

\section*{Acknowledgments}
We would like to thank all the anonymous reviewers for their insightful comments. The work was also supported by National Key R\&D Program of China (No. 2022YFB3103900), Natural Science Foundation of Guangdong Province (Project No. 2023A1515011959), Shenzhen Basic Research (General Project No. JCYJ20220531095214031), Shenzhen International Cooperation Project (No. GJHZ20220913143 008015), and the Major Key Project of PCL (Grant No.PCL2022A03).

\normalem
\bibliographystyle{IEEEtran}
\bibliography{ref}

\end{document}

%% file: sections/0_abs.tex
Automatic API method recommendation is an essential task of code intelligence, which aims to suggest suitable APIs for programming queries. 
Existing approaches can be categorized into two primary groups: retrieval-based and learning-based approaches. Although these approaches have achieved remarkable
success, they still come with notable limitations. The retrieval-based approaches rely on the text representation capabilities of embedding models, while the learning-based approaches require extensive task-specific labeled data for training.
To mitigate the limitations,
we propose {\tool}, a generative API recommendation approach through enhanced in-context learning (ICL). {\tool} has a powerful representation capability and can make effective recommendations with only a few examples via ICL. To overcome the limitations of standard ICL in capturing task-specific knowledge, {\tool} involves two main
components: 
(1) Diverse Examples Selection. {\tool} searches for similar posts to the programming queries from the lexical, syntactical, and semantic perspectives, providing more informative 
examples for ICL.
(2) Guided API Recommendation. {\tool} enables
large language models (LLMs) to perform reasoning before generating API recommendations, where the reasoning involves fine-grained matching between the 
task intent behind the queries and the factual knowledge of the APIs. With the reasoning process, {\tool} makes recommended APIs better meet the programming requirement of queries and also enhances the interpretability of results.
We compare {\tool} with four existing approaches on two publicly available benchmarks. Experiments show that {\tool} outperforms the best baseline CLEAR by 105.8\% in method-level API recommendation and 54.3\% in class-level API recommendation in terms of SuccessRate@1. 
Besides, {\tool} achieves an average 49.87\% increase compared to the zero-shot performance of popular LLMs such as GPT-4
in method-level API recommendation regarding the SuccessRate@3 metric.

%% file: sections/1_intro.tex
Application Programming Interfaces (APIs) play an important role in modern software development. They enable developers to access and leverage external functionalities, services, and resources, which can enhance the efficiency of application development. 
However, the rapid evolution of API libraries and services~\cite{DBLP:conf/wcre/HouY11, DBLP:journals/tse/YuBSM21, DBLP:journals/tse/ChenGRP0L23} poses a great challenge to developers: how to select the most appropriate APIs for their specific programming requirements. To tackle this challenge, various automated API method recommendation approaches have been proposed~\cite{DBLP:conf/icse/McMillanGPXF11, DBLP:conf/fase/ZhangZL11, DBLP:conf/sigsoft/ChanCL12, DBLP:conf/wcre/RahmanRL16, DBLP:conf/kbse/HuangXXLW18, DBLP:conf/icse/WeiHH0022, DBLP:conf/sigsoft/GuZZK16,DBLP:journals/jcst/LingZLX19, DBLP:journals/tse/ZhouYCHMG22}. For a programming task, they recommend some suitable APIs by evaluating the similarity between the natural language description of this task and the functional descriptions of APIs or checking if these APIs have been applied to similar tasks.

\begin{figure}
    \centering
    \includegraphics[scale=0.45]{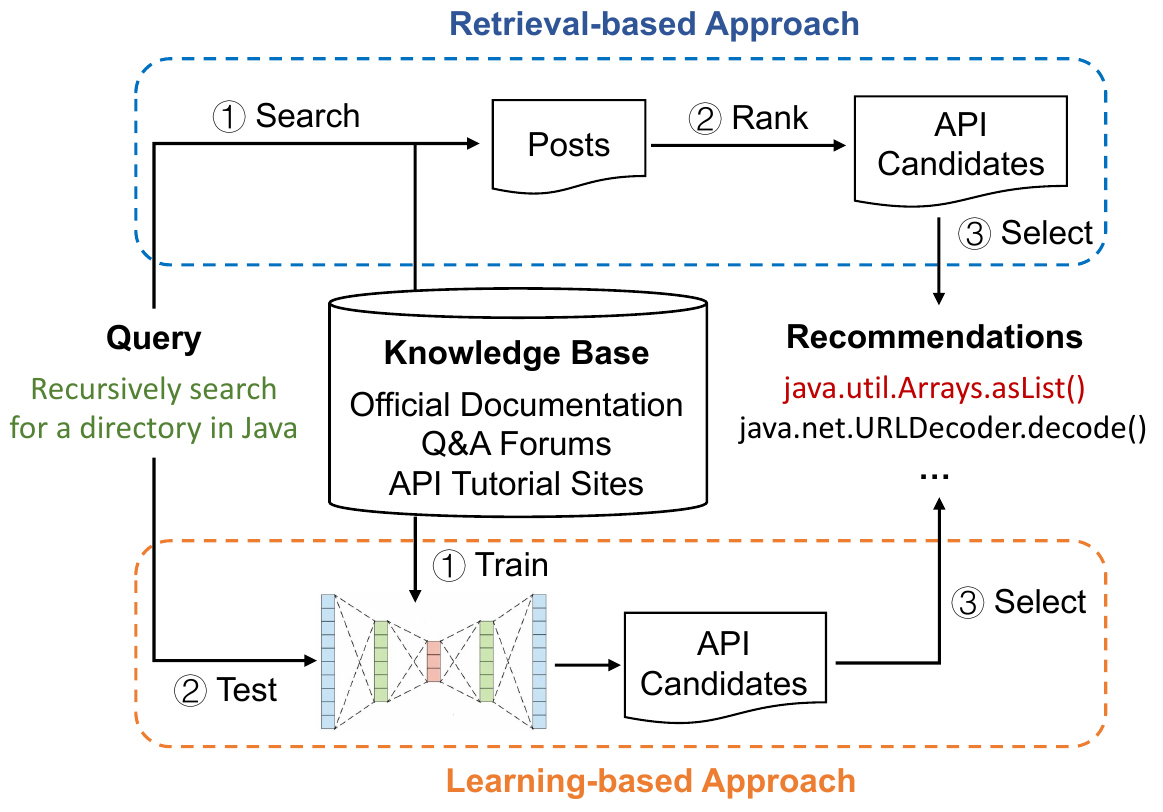}
    \caption{Two types of API recommendation approaches.}
    \label{fig:category}
\end{figure}

Existing approaches can be categorized into two primary groups: retrieval-based approaches (e.g., RACK~\cite{DBLP:conf/wcre/RahmanRL16}, BIKER~\cite{DBLP:conf/kbse/HuangXXLW18} and CLEAR~\cite{DBLP:conf/icse/WeiHH0022}) and learning-based approaches (e.g., DeepAPI~\cite{DBLP:conf/sigsoft/GuZZK16}). 
The typical workflow for these two types of approaches is illustrated in Fig.~\ref{fig:category}. They rely on a knowledge base that includes all the known APIs. This knowledge base contains official documentation with detailed API functional descriptions, programming-related Q\&A forums like Stack Overflow~\cite{StackOver}, and API tutorial websites like GeeksforGeeks~\cite{Geeks4geeks}. 
For retrieval-based approaches, they first search for relevant posts from the knowledge base. These posts are then ranked by measuring their similarity to the query, which helps in identifying API candidates. The top-k candidates are finally selected as the recommended APIs. The state-of-the-art retrieve-based approach is CLEAR, which employs the BERT sentence embedding model~\cite{DBLP:conf/naacl/DevlinCLT19} to capture
sequential semantic information of queries. It also uses contrastive training~\cite{DBLP:journals/corr/abs-1807-03748} to improve the understanding of query semantics. 
These retrieval-based methods mainly rely on evaluating the similarity between various texts, like queries and posts, to provide recommendations. Thus, their effectiveness is limited by the
representation capabilities of the
embedding models. 
For example, given a real-world query ``\textit{Recursively search for a directory in Java}''~\cite{question1} as shown in Fig.~\ref{fig:category}, CLEAR first identifies the most similar post ``\textit{find annotated Methods Recursively}''~\cite{question2} from the knowledge base, as this post is syntactically and lexically close to the query. Based on the answer from the post, CLEAR recommends the top API ``\textit{java.util.Arrays.asList()}''. However, this API fails to
solve the given programming query.
For learning-based approaches, they utilize deep learning techniques to discover the relationships between queries and APIs.
The first learning-based approach is DeepAPI, which models API recommendation task as a machine translation problem. DeepAPI uses a Recurrent Neural Network (RNN) Encoder-Decoder model~\cite{DBLP:conf/emnlp/ChoMGBBSB14} to encode a given query into a fixed-length context vector and generate an API sequence based on this context vector. 
These learning-based approaches are limited by insufficient training data in this task domain. 
They require a large number of query-API pairs for training, however, the current training data are hard to cover all the APIs. For example, Stack Overflow contains only about 12,000 API-related posts~\cite{DBLP:journals/tse/PengLGLWGL23}, which are far fewer than the number of APIs (i.e., 30,000) from the official documentation.
Given the same query ``\textit{Recursively search for a directory in Java}'', DeepAPI generates an API sequence \{\textit{``File.isDirectory'', ``File.getDescent'', ``File.getAbsolutePath''}\}, which also does not include the correct API.

\noindent \textbf{Our work.} In this paper, we propose {\tool}, the first generative API method recommendation approach based on enhanced in-context learning (ICL). 
Benefiting from the extensive text encoded in large language models (LLMs), {\tool} has a powerful representation capability, which makes it better understand queries and API documentation. Furthermore, {\tool} can make effective recommendations with only a few examples via ICL, without requiring a large amount of labeled data. 
To overcome the limitations of standard ICL in capturing task-specific knowledge, {\tool} 
involves two main components: diverse example selection and guided API recommendation. 
These components include three main phases: example retrieval, prompt construction, and API recommendation.
In the example retrieval phase, {\tool} uses the given query to search for relevant posts from an API-related posts corpus. These relevant posts serve as demonstration examples for the following phases, where each post contains a programming question and its corresponding API answer.
In the prompt construction phase, {\tool} first extracts the task intent behind questions by analyzing the constituency tree and grammatical elements, and then detects factual knowledge about APIs based on a constructed official description dictionary. 
Next, {\tool} performs a fine-grained matching between the task intent and the factual knowledge to generate the reasoning prompt.
The reasoning prompt provides guidance to LLMs on how to analyze queries and recommend appropriate APIs effectively. 
In the last API recommendation phase, {\tool} combines questions, reasoning prompts, and API answers to create a demonstration. Using the demonstration and the given query as an input prompt, {\tool} leverages the LLM to generate API methods and their corresponding reasons.

We evaluate {\tool} using two widely-used API recommendation datasets provided by Huang \etal\cite{DBLP:conf/kbse/HuangXXLW18}, comprising 33,000 Java questions, and collected by Peng \etal\cite{DBLP:journals/tse/PengLGLWGL23}, including 6,563 Java questions. 
Our experimental results demonstrate that {\tool} outperforms the state-of-the-art baseline CLEAR~\cite{DBLP:conf/icse/WeiHH0022} in method-level API recommendation, achieving improvements of 61.29\%, 82.61\%, 72\% and 28.26\% in terms of SuccessRate@3, MAP@3, MRR and NDCG@3, respectively. 
Through ablation experiments, we find that adding retrieved examples enhances {\tool} by 42.2\%, and introducing reasoning prompts further improves {\tool} by 79.7\% in terms of SuccessRate@1. The source code is publicly accessible at \url{https://github.com/hitCoderr/APIGen}.

\noindent \textbf{Contributions.}
In summary, our main contributions in this paper are as follows:

\begin{itemize}
\item To the best of our knowledge,  we are the first work to propose a generative API recommendation approach based on enhanced in-context learning, named {\tool}.
\item We propose a novel reasoning prompt to incorporate fine-grained matching between the query's task intent and the API's factual knowledge into large language models, making them better understand queries and generate more suitable API recommendation.
\item We conduct extensive experiments to evaluate {\tool} on two benchmark datasets. The experimental results demonstrate that {\tool} substantially improves the performance of the prior approaches on both method-level and class-level API recommendation.
\end{itemize}

\noindent \textbf{Outline.} The rest of the paper is organized as follows: Section~\ref{sec:background} provides an overview of the study's background. Section~\ref{sec:approach} presents the architecture of the proposed {\tool}. Section~\ref{sec:setup} and Section~\ref{sec:results} detail the experimental setup and present the results, respectively. Section~\ref{sec:discussion} presents a case study and analyses potential threats to validity. Section~\ref{sec:related} briefly describes the related works. In the end, in Section~\ref{sec:conclusion}, we summarize the whole work. 

%% file: sections/2_background.tex
\subsection{Large Language Models}

Large Language Models (LLMs) have become a ubiquitous part of Natural Language Processing (NLP) due to their remarkably exceptional performance~\cite{DBLP:conf/nips/BrownMRSKDNSSAA20,DBLP:journals/corr/abs-2303-08774}. 
These models are trained on a massive textual corpus~\cite{DBLP:journals/jmlr/RaffelSRLNMZLL20,DBLP:journals/tse/MillerS76,DBLP:conf/nips/Ouyang0JAWMZASR22,DBLP:conf/nips/VaswaniSPUJGKP17} using self-supervised objectives such as Masked Language Modeling~\cite{DBLP:conf/emnlp/FengGTDFGS0LJZ20} and Causal Language Modeling~\cite{DBLP:conf/iclr/NijkampPHTWZSX23}.
Most LLMs follow the Transformer architecture \cite{DBLP:conf/nips/VaswaniSPUJGKP17}, which contains an encoder for input representation and a decoder for output generation.
To date, LLMs have been applied to various domains and achieved great success ~\cite{DBLP:conf/nips/BrownMRSKDNSSAA20,DBLP:conf/kbse/HuangYXX0022,DBLP:conf/icse/WangJLYX0L22}.

The size of LLMs has increased significantly in the past few
years. For example, the parameters of recent LLMs like GPT-3~\cite{DBLP:journals/jmlr/RaffelSRLNMZLL20} and PALM-E \cite{DBLP:conf/icml/DriessXSLCIWTVY23} are over one hundred billion. 
In addition, there are also LLMs with billion-level parameters trained for some specific tasks, such as code generation, code completion, and code summarization~\cite{DBLP:journals/corr/abs-2309-07062,DBLP:journals/corr/abs-2308-12950,DBLP:conf/icse-apr/PrennerBR22,DBLP:journals/corr/abs-2203-07814}. 
Particularly, OpenAI’s Codex~\cite{DBLP:conf/icse-apr/PrennerBR22} is a large pre-trained code model that is capable of powering Copilot, and AlphaCode~\cite{DBLP:journals/corr/abs-2203-07814} is a 41-billion-large model trained for generating code in programming competitions like Codeforces.
Recently, LLMs like ChatGPT~\cite{chatgpt2022} and GPT-4~\cite{DBLP:journals/corr/abs-2303-08774} have
also shown impressive performance in many code intelligence tasks.
%
%
Prompt engineering techniques have been proposed to improve the performance of LLMs on specific tasks by carefully designing the input prompt. Among prompt engineering techniques, Chain-of-Thought (CoT) has been shown to elicit stronger reasoning from LLMs by asking the model to incorporate intermediate reasoning steps when solving a problem~\cite{DBLP:journals/corr/abs-2210-06726,DBLP:journals/corr/abs-2206-02336,DBLP:journals/corr/abs-2207-00747}. 
In API recommendation task, using LLMs directly presents two problems. Firstly, understanding the specific intentions behind diverse programming requirements is difficult for LLMs. Besides, LLMs are often seen as ``black-box'' models, making it hard for users to understand the reasoning behind recommended APIs.

\subsection{In-context Learning}

\begin{figure}
    \centering
    \includegraphics[scale=0.55]{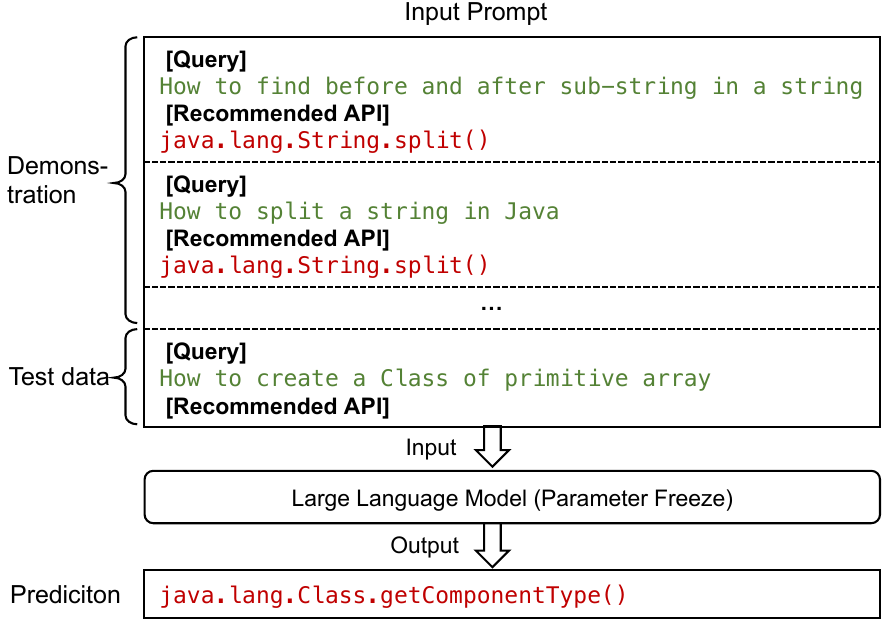}
    \caption{An illustration of standard ICL on API recommendation.}
    \label{fig:icl}
\end{figure}

As the size of LLMs continues to increase, tuning a LLM for downstream tasks can be expensive and impractical for researchers.
To alleviate this issue, in-context learning (ICL) leverages a demonstration in the prompt to help the model learn the input-output mapping of the downstream tasks without requiring parameter updates~\cite{DBLP:journals/jmlr/RaffelSRLNMZLL20,DBLP:journals/corr/abs-2301-00234}.
This new paradigm has achieved impressive results in various tasks such as logic reasoning and program repair~\cite{DBLP:conf/nips/Wei0SBIXCLZ22,DBLP:conf/icse/XiaWZ23}.

ICL is a method used with LLMs to help them understand and respond to specific tasks. The core idea is to use analogy-based learning: LLMs can understand the task better and generate more accurate results by providing a suitable demonstration. For example, as shown in Fig.~\ref{fig:icl}, to employ LLMs recommend APIs for a query, we first provide a demonstration with two examples of the query and its corresponding API answer, and then LLMs can identify patterns from the provided context and make the prediction. Clearly, ICL not only can help LLMs understand tasks better but also offer an interpretable way to interact with LLMs. Moreover, ICL eliminates the need for extensive training, making the process efficient.

%% file: sections/3_approach.tex
\begin{figure*}
    \centering
    \includegraphics[scale=0.46]{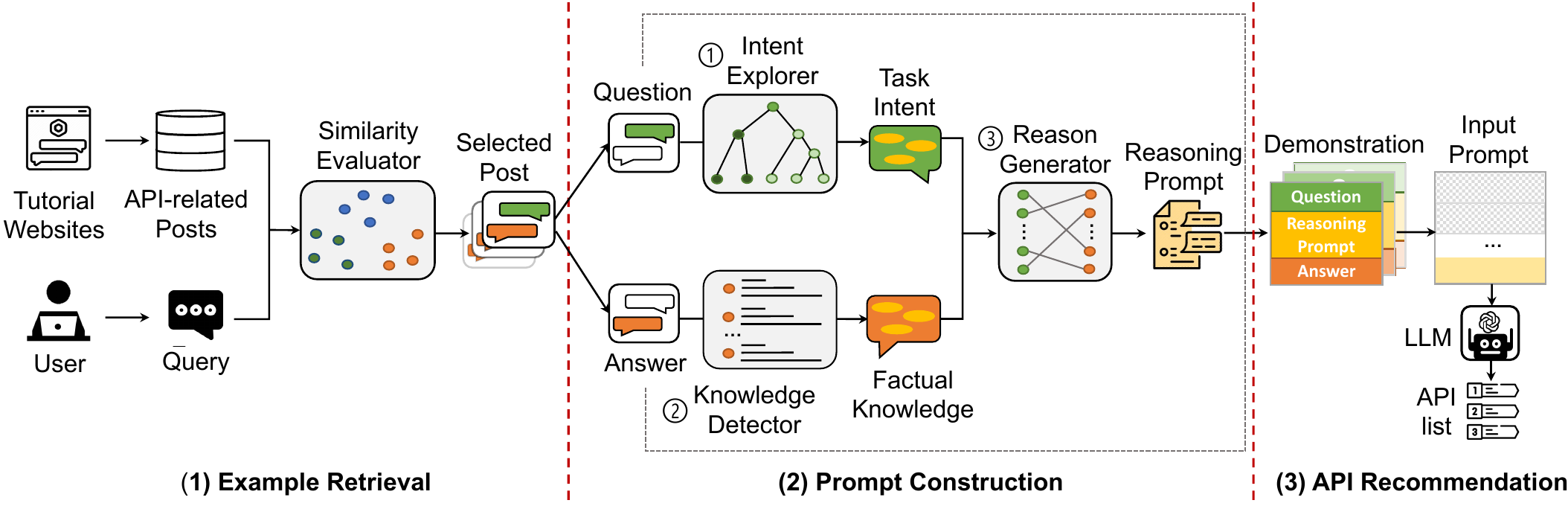}
    \caption{The overview of {\tool}.}
    \label{fig:framework}
\end{figure*}

In this section, we propose a generative API method recommendation approach via enhanced in-context learning, named {\tool}. 
We first present the overview of {\tool} and then describe its details in the following subsections.

\subsection{Overview}
Fig.~\ref{fig:framework} shows the overview framework of {\tool}, which sequentially performs the following three steps to generate suitable APIs for an input query: 
\begin{enumerate}
    \item {\it Example Retrieval.} Searching for relevant posts from the API-related posts corpus as demonstration examples for ICL. 
    These examples provide different real-world programming queries and API answers, helping {\tool} to learn which APIs should be selected as the solution for a specific development problem.
    \item {\it Prompt Construction.}
    Generating the reasoning prompt by performing a fine-grained matching between the task intent behind the question and the factual knowledge of the API answer. The reasoning prompt guides LLMs on how to analyze queries and recommend appropriate APIs effectively.  
    \item {\it API Recommendation.} Combining the demonstration with the given query as an input prompt for LLMs, where the demonstration is formed by $<$question, reasoning prompt, answer$>$. Using the input prompt, LLMs generate APIs and corresponding explanations. The guided API recommendation makes the predicted APIs better meet the programming requirement of queries and also enhances the interpretability of results.
     
\end{enumerate}

\subsection{Example Retrieval} \label{sec:exampleRetrieval}

This phase aims to select relevant posts from an API-related post set, which can be obtained from Q\&A forums and tutorial websites like Stack Overflow~\cite{StackOver}, GeeksforGeeks~\cite{Geeks4geeks}, Java2s~\cite{Java2s} and Kode Java~\cite{Kode}. 
Following the prior works~\cite{DBLP:conf/icse/WeiHH0022, DBLP:conf/kbse/HuangXXLW18, DBLP:conf/wcre/RahmanRL16}, we consider a post relevant to the query if this post's question is semantically similar to this query. Here, we design a \textit{similarity evaluator} to filter out the relevant posts.

In \textit{similarity evaluator}, we employ three retrieval-based techniques, including BM-25~\cite{DBLP:journals/ftir/RobertsonZ09}, SBERT~\cite{DBLP:conf/emnlp/ReimersG19} and CodeT5~\cite{wang-etal-2021-codet5}. Intuitively, we separately use the three models to capture a comprehensive understanding of sentences from different perspectives. 
Specifically, BM-25 measures the lexical similarity between two sentences and evaluates them at the word-choice level.
In contrast to BM-25, which focuses on individual words, SBERT captures the overall semantics of entire sentences.
On the other hand, CodeT5 is a pre-trained model for programming tasks and easily understands natural language descriptions of coding tasks.
By leveraging the above three models, we can more accurately measure the relevance between the input query and the questions of the posts.
Using the \textit{similarity evaluator}, we retrieve the top-$n$ posts ($<$question, answer$>$) as examples.

\subsection{Prompt Construction} \label{sec:prompt}

\begin{figure}[h]
    \centering
    \includegraphics[scale=0.26]{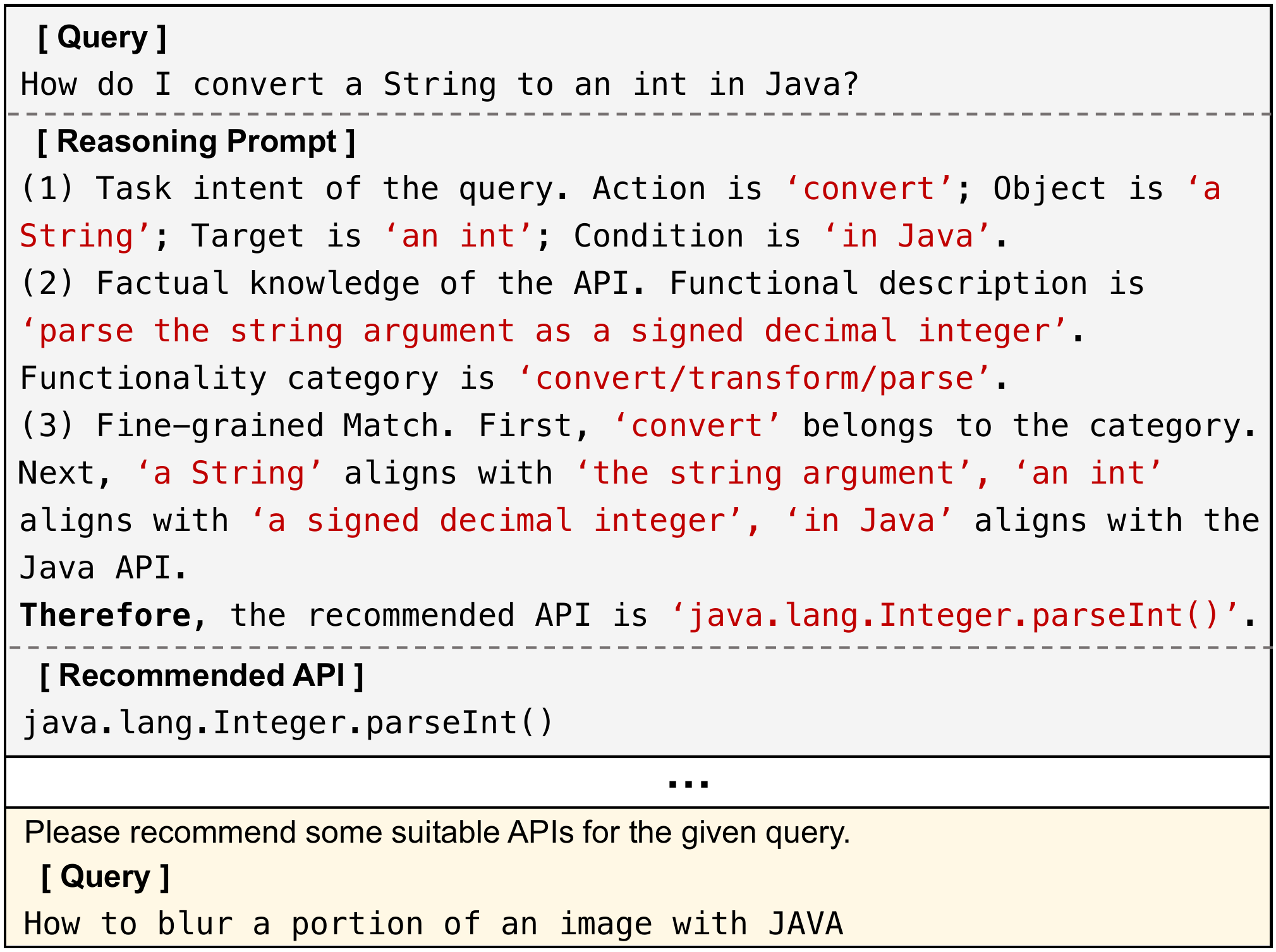}
    \caption{An example of input prompt in {\tool}.}
    \label{fig:example}
\end{figure}

This phase aims to create the reasoning prompt, which explains why certain APIs are selected as the answer, providing guidance to LLMs on generating suitable APIs. 
Creating the reasoning prompt involves three steps: 1) analyzing the task intent behind the question using  {\it intent explorer} 2) obtaining the factual knowledge of the answer via  {\it knowledge detector} 3) performing a fine-grained matching between the obtained task intent and factual knowledge via \textit{reason generator}.


\subsubsection{Intent Explorer} \label{subsubsection:intent}
For a given question, the \textit{intent explorer} performs an analysis of its intent, including three steps:
\begin{itemize}
    \item {\it Question Refinement.} In real-world scenarios, questions often contain non-programming related information or lack crucial keywords. 
    We refine them via an LLM (GPT3.5) to either distill the core content or fill in the missing parts. 
    This is achieved through the prompt: ``If the sentence lacks a verb, add an appropriate one; otherwise, extract the main content extracting their core content.''. 
    Taking the question ``\textit{How do I convert a String to an int in Java}'' in Fig.~\ref{fig:example} as an example, we distill its essential part: ``\textit{convert a String to an int in Java}''. 
    Additionally, for the incomplete question \textit{16-bit hex string to signed int in Java}, we can insert a verb \textit{convert}.
    \item {\it Question Classification.} After reformulating the question, we leverage AllenNLP~\cite{DBLP:journals/corr/abs-1803-07640}, a widely-used natural language processing tool, to classify its content. First, we parse the question into a constituency tree to identify syntactic categories: Verb (VB), Sentence (S), Noun Phrase (NP), Verb Phrase (VP), and Prepositional Phrase (PP). By doing so, we typically classify a question into one of three structural forms: VB+NP+(PP/S), VB+NP+PP+(PP/S), and VB+S. 
    Next, we analyze the grammatical elements of a question and assign them to one of six syntactic roles by part-of-speech tagging. 
    These roles include verb, direct object, preposition, preposition object, direct object's modifier, and preposition object's modifier. 
    \item {\it Question Deconstruction.} To acquire the intent from the reformulated question, we divide it into four key components: 
    \textbf{Action} indicates what function it requires to perform, which is the core operation of the question, such as \textit{get}, \textit{convert}, or \textit{create}. 
    \textbf{Object} indicates what it operates on, which is the primary entity of the question, such as data types, libraries, or frameworks. 
    \textbf{Target} indicates what it wants to achieve, which is the result of the question.
    \textbf{Condition} indicates any rules or restrictions associated with the question such as specific programming languages (e.g., \textit{in Java}) or data formats (e.g., \textit{CSV-format input data}). 
    By breaking the question down into these four parts, we can better understand and extract its essential details, ensuring more accurate API recommendations. Table~\ref{tab:question_intent} shows how to derive the above four parts based on the constituent forms and syntactic roles. 
\end{itemize}
Taking the question ``\textit{How do I convert a String to an int in Java?}'' in Fig.~\ref{fig:example} as an example, we first reformulate it as ``\textit{convert a String to an int in Java}''. After parsing, we classify it as having the structure ``VB+NP+PP+PP'', with six syntactic roles assigned as follows: the verb is ``convert'', the direct object (dobj) is ``String'', the direct object's modifier (dmod) is ``a'', the preposition object (pobj) is ``int'' and the preposition object's modifier (pmod) is ``an''. Referring to the rules in Table~\ref{tab:question_intent}, we can extract the question's intent as follows: the action is ``convert'', the object is ``a String'', the target is ``an int'', and the condition is ``in Java''.

\begin{table}
\centering
\renewcommand\arraystretch{1.25}
\caption{Intent exploration rules based on constituency tree and syntactic roles. ``dobj'' indicates ``direct object'', ``pobj'' indicates ``preposition object'', ``dmod'' indicates direct object's modifier, ``pmod'' indicates ``preposition object's modifier''.}
\label{tab:question_intent}
    \begin{tabular}{c|c|c|c|c}
     \bottomrule
      Constituency & Action & Object & Target & Condition \\
     \hline
      VB+NP+(PP/S)  & verb & N/A & dobj+dmod & PP/S \\
     \hline
      VB+NP+PP+(PP/S)  & verb & dobj+dmod & pobj+pmod & PP/S \\
     \hline
      VB+S & verb & N/A & N/A & S \\
    \toprule
    \end{tabular}
\end{table}

\subsubsection{Knowledge Detector}
To gather factual knowledge of APIs, we establish a Java API dictionary with 30,287 method-description pairs. The dictionary is built by parsing the JDK 1.8 API reference documentation~\cite{javadoc} and extracting all API methods from the HTML file of each class. 
It is noted that our dictionary does not include the deprecated methods, such as the method described by ``\textit{Deprecated. use Socket.getOption (SocketOption) instead}''.
Descriptions in the dictionary can be divided into 87 functionality categories~\cite{DBLP:conf/sigsoft/Xie0LTXZZ20} based on the meaning of its verb. 
For example, consider the description ``\textit{parse the string argument as a signed decimal integer}'' in Fig.~\ref{fig:example}. It contains the verb ``parse'', which means ``transform something into other forms'', thus falls into the ``convert/transform/parse'' category. 
We annotate the descriptions using a fine-tuned BERT model provided by~\cite{DBLP:conf/sigsoft/Xie0LTXZZ20}, and append the functionality categories of them in the API dictionary. 
Through the dictionary, the {\it knowledge detector} can retrieve the functional description and functionality category of the API answer. 
 

\subsubsection{Reason Generator}

We perform fine-grained matching between the task intent of the question and the factual knowledge of the answer to generate a reasoning prompt. 
Specifically, we first match the action in the intent with the functionality category in the knowledge, and then align the respective entities in the intent and knowledge based on their semantic roles, which include object, target, and condition.
We provide a template for generating the reasoning prompt, as shown in Fig.~\ref{fig:example}. In this example, we first obtain the intent of the question and factual knowledge of the API answer. 
Next, we perform fine-grained matching between them. The alignment provides a clear explanation of why the recommended API can meet the programming requirement of the query.

\subsection{API Recommendation}

This phase aims to guide LLMs in generating high-quality API recommendations. First, we combine all questions, reasoning prompts, and API answers to create a demonstration. Next, we feed the demonstration and the given query as an input prompt to LLMs, as presented in Fig.~\ref{fig:example}.
By learning from the demonstrations within the input prompt, LLMs generate both API recommendation and corresponding reasoning processes. 
The guided API recommendation makes the prediction better meet the programming requirement of queries and enhances the interpretability of results.



%% file: sections/4_setup.tex
\subsection{Research Questions} \label{sec:RQ}

We conduct extensive experiments to evaluate the proposed approach with the aim of answering the following research questions:

\begin{itemize}[leftmargin=*]

\item {\textbf{RQ1:}} How effective is {\tool} compared with the state-of-the-art API recommendation approaches?
\item {\textbf{RQ2:}} What are the impacts of two main modules (i.e., \textit{Example Retrieval} and \textit{Prompt Construction}) in {\tool}?
\item {\textbf{RQ3:}} What are the effects of using different examples in {\tool}?
\item {\textbf{RQ4:}} How does {\tool} perform on different large language models?

\end{itemize}

\subsection{Baselines} \label{sec:baseline}

We choose the following four API recommendation approaches as our baselines:

\begin{itemize}[leftmargin=*]

\item {\textbf{RACK~\cite{DBLP:conf/wcre/RahmanRL16}}} recommends APIs by searching the relevant API from Stack Overflow.
It first builds a keyword-API database from Stack Overflow questions and answers. Then, for a given query, RACK ranks API classes based on keyword similarity with the query.

\item {\textbf{DeepAPI~\cite{DBLP:conf/sigsoft/GuZZK16}}} models API recommendation task as a machine translation problem. 
It uses a Recurrent Neural Network (RNN) Encoder-Decoder model to encode a given query into a fixed-length context vector and generate an API sequence based on this vector. 

\item {\textbf{BIKER~\cite{DBLP:conf/kbse/HuangXXLW18}}} trains a word embedding model to calculate the similarity between a given query and the Stack Overflow posts. It then selects the top-N API answers from these posts as candidates. 
These top-N answers are further refined by comparing the query's similarity to official API documentation descriptions.

\item {\textbf{CLEAR~\cite{DBLP:conf/icse/WeiHH0022}}} is based on BERT sentence embedding~\cite{DBLP:conf/naacl/DevlinCLT19} and contrastive learning~\cite{DBLP:journals/corr/abs-1807-03748}. Given a query, CLEAR first selects a set of candidate Stack Overflow posts based on BERT sentence embedding similarity and re-ranks them using a BERT-based classification model to recommend the top-n APIs.

\end{itemize}

\subsection{Datasets} \label{sec:dataset}
To comprehensively evaluate the performance of {\tool}, we adopt two widely used datasets: APIBENCH-Q~\cite{DBLP:journals/tse/PengLGLWGL23} and BIKER-Dataset~\cite{DBLP:conf/kbse/HuangXXLW18}. The details of two datasets are described as follows.

\begin{itemize}[leftmargin=*]

\item {\textbf{APIBENCH-Q}} contains 6,563 Java questions sourced from well-known platforms, including Stack Overflow and tutorial websites. From APIBENCH-Q, we randomly select 500 questions with API answers to create a test set, while the remaining questions are utilized as a training set.

\item {\textbf{BIKER-Dataset}} comprises 33,000 Java-related questions extracted from the official data dump of Stack Overflow, which is provided by BIKER and used as the training set.  
In addition, BIKER also provides a test dataset~\cite{DBLP:conf/kbse/HuangXXLW18, DBLP:conf/icse/WeiHH0022}, which includes 413 manually selected and verified SO questions with corresponding API answers. 
\end{itemize}

We train the baseline models and select examples for {\tool} using the training sets. And we evaluate the performance of both {\tool} and the baseline models on the testing sets. 

\subsection{Metrics}

We adopt four widely-used metrics ~\cite{DBLP:conf/sigsoft/GuZZK16, DBLP:conf/icse/WeiHH0022, DBLP:journals/tse/PengLGLWGL23}: Success Rate, Mean Reciprocal Rank (MRR), Mean Average Precision (MAP), and Normalized Discounted Cumulative Gain (NDCG) to evaluate the performance of {\tool} and other baselines. 

\begin{itemize}[leftmargin=*]

\item {$Success Rate@k$} evaluates the ability of a model to recommend correct APIs based on the top-k returned results regardless of the orders.

\begin{equation}
    Success Rate@k = \frac{\sum_{i=1}^N HasCorrect_k(q_i)} {N}
\end{equation}

where $HasCorrect_k(q)$ returns 1 if the top-k results of query $q$ contain the correct API, otherwise it returns 0.

\item {$MAP@k$} measures the effort needed to find the first correct answer in the recommended list. 

\begin{equation}
\begin{aligned}
    AveP@k &=  \frac{\sum_{i=1}^N P_k(i) \times rel(i)} {m} \\
    MAP@k &= \frac{\sum_{i=1}^N Ave P_k (q_i)} {N}
\end{aligned}
\end{equation}

where $rel(i)$ returns 1 if the $i_{th}$ result is the correct API, otherwise it returns 0.

\item {$MRR$} considers the ranks of all correct answers.

\begin{equation}
     MRR = \frac{\sum_{i=1}^N 1/ firstpos (q_i)} {N}
\end{equation}

where $firstpos(q)$ returns the position of the first correct API in the results, if it cannot find a correct API in results, it returns $+\infty$.

\item {$NDCG@k$} measures the quality of the recommended list by considering the relevance score for each position in the list.

\begin{equation}
\begin{aligned}
    DCG@k &= \sum_{t=1}^k \frac{rel_t (q_i)} {log_2(t+1)} \\
    NDCG@k &= \frac{\sum_{i=1}^N \frac{DCG@k(q_i)}{IDCG@k(q_i)}} {N}
\end{aligned}
\end{equation}

where $rel_t (q)$ returns 2 if $t_{th}$ result exactly matches one correct API, and it returns 1 if $t_{th}$ result matches the API class but fails to match the API, otherwise it returns 0. $IDCG@k$ is the best $DCG@k$ by re-arranging the order of current results.

\end{itemize}

\subsection{Implementation Details}

For the four API recommendation baselines RACK~\cite{DBLP:conf/wcre/RahmanRL16}, DeepAPI~\cite{DBLP:conf/sigsoft/GuZZK16}, BIKER~\cite{DBLP:conf/kbse/HuangXXLW18} and CLEAR~\cite{DBLP:conf/icse/WeiHH0022}, we directly use the replication packages released by the authors and other researchers. For the implementation of {\tool}, We utilize the GPT-3.5 (text-davinci-003)~\cite{DBLP:conf/nips/BrownMRSKDNSSAA20} in our paper for all experiments in the first three RQs. In RQ4, we further use the API of ChatGPT (gpt-3.5-turbo)~\cite{chatgpt2022} and GPT-4 (gpt-4)~\cite{DBLP:journals/corr/abs-2303-08774} for experiments. As for the hyperparameters of the APIs, we set temperature to 0.15, max generation length to 512, sampling number to 5, and adopt nuclear sampling~\cite{DBLP:conf/iclr/HoltzmanBDFC20} with top-p set as
0.95. Besides, we set the number of retrieved examples to 3 and the example selection method to SBERT by default.
We conduct all the experiments on a server with four NVIDIA Tesla V100 GPUs.

%% file: sections/5_result.tex
This section presents our experiment results and answers for the four research questions in Section~\ref{sec:RQ}.

\subsection{Effectiveness of {\tool} (RQ1)}

\input{sections/5.0_main_table}

\noindent \textbf{Experimental Design.} To answer this research question, we compare {\tool} with the baselines listed in Section~\ref{sec:baseline} on the two datasets listed in Section~\ref{sec:dataset}. Note that, we exclude RACK in this research question as it recommends API at class-level only.

\noindent \textbf{Results.} Table~\ref{tab:comparison_method} and Table~\ref{tab:comparison_class} show the results of {\tool} compared with the baselines in the method-level and class-level API recommendation, respectively. We have also conducted the Wilcoxon signed-rank test~\cite{wilcoxon1992individual} (\textit{p}-value$<$0.01) to compare the performance of {\tool} and baselines. The test result suggests that {\tool} achieves significantly better performance than all the baselines.

\noindent \textbf{Analyses.}
\uline{(1) {\tool} achieves higher accuracy at both method-level and class-level.} 
By analyzing the top-5 method-level recommendation results in Table~\ref{tab:comparison_method}, we observe that {\tool} outperforms the best approach CLEAR by 37.5\% in terms of Success Rate on APIBENCH-Q. This improvement of {\tool} over CLEAR is even more significant for top-1 recommendation, where {\tool} outperforms CLEAR by 105.88\% in terms of Success Rate on APIBENCH-Q. The SuccessRate@1 measures the capability to correctly predict the API in the first position. In the class-level recommendation, all the approaches demonstrate an obvious improvement. For example, as shown in Table~\ref{tab:comparison_class}, the SuccessRate@1 of BIKER has reached 0.33, marking an
175\% increase compared to the method-level recommendation on APIBENCH-Q. According to
the top-1,3,5 results, {\tool} achieves a consistent improvement of 6.34\% $\sim$ 54.29\% on Success Rate compared to CLEAR on APIBENCH-Q. Furthermore, {\tool} achieves the highest SuccessRate@5, with a score of 0.95. This indicates that {\tool} can find the correct API class within the top-5 returned results for nearly all queries in BIKER-Dataset. These results demonstrate that the generative API recommendation approach {\tool} is highly effective.
\uline{(2) {\tool} ranks the correct APIs better.} By analyzing the metrics for evaluating API ranking in Table~\ref{tab:comparison_method}, such as MAP@k, MRR and NDCG@k, we observe that {\tool} outperforms the best approach CLEAR by 72\% on MAP@5, 72\% on MRR and 17.65\% on NDCG@5 in APIBENCH-Q. This indicates that {\tool} not only has a greater ability to recommend the correct APIs but also effectively ranks them ahead in the returned results. 

\begin{tcolorbox}[breakable,width=\linewidth-2pt,boxrule=0pt,top=3pt, bottom=3pt, left=4pt,right=4pt, colback=gray!15,colframe=gray!15]
\textbf{Answer to RQ1:} 
{\tool} outperforms the best baseline CLEAR on both datasets, with particularly notable improvements of 105.88\% and 54.29\% in method-level and class-level recommendation, respectively, in terms of SuccessRate@1.
\end{tcolorbox}

\subsection{Impacts of Different Modules in {\tool} (RQ2)}

\input{sections/5.1_aba_table}

\input{sections/5.2_para_figure}

\noindent \textbf{Experimental Design.} To answer this research question, we perform ablation studies by considering the following two variants of {\tool}.

\begin{itemize}[leftmargin=*]

\item {$\text{\tool}_{\text{w/o\ example}}$:} In this variant, we exclude any retrieved posts, relying solely on the given query as the input prompt.
 
\item {$\text{\tool}_{\text{w/o\ reasoning}}$:} In this variant, we provide retrieved posts but exclude reasoning prompts as the input prompt.

\end{itemize}

\noindent \textbf{Results.} Table~\ref{tab:aba} shows the results of {\tool} compared with two variants in the method-level and class-level API recommendation. We choose Success Rate and MAP as representative metrics to evaluate the accuracy and ranking performance of APIGen. 

\noindent \textbf{Analyses.} \uline{Both modules are essential for {\tool} to achieve optimal performance.} 
Experimental results reveal that removing the example retrieval module leads to a large drop in {\tool}'s performance. For example, on the APIBENCH-Q, the overall Success Rate and MAP on method-level API recommendation decrease by 34.14\% and 40.07\%, respectively, while the performance on class-level API recommendation experiences a 44.28\% and 57.33\% drop in overall Success Rate and MAP, respectively.
This is because without the example retrieval module, {\tool} loses the context provided by existing examples, which is crucial for understanding query context and identifying relevant APIs. 
Besides, removing the reasoning prompt and solely providing examples to the LLM results in a slight performance decrease in {\tool}. For example, in method-level API recommendation, the average SuccessRate@1 and MAP@1 decrease by 11.55\% and 10.20\% on both datasets, respectively. The drop in performance can be attributed
to two main factors: lacking of interpretative and reasoning abilities. 
The reasoning prompt construction module plays a crucial role in helping {\tool} better understand query intent and infer potential APIs. It guides the model's reasoning process, explaining why a particular API is selected as the answer. The decline highlights the significance of interpretative and reasoning abilities for API recommendation.

\begin{tcolorbox}[breakable,width=\linewidth-2pt,boxrule=0pt,top=1pt, bottom=1pt, left=4pt,right=4pt, colback=gray!15,colframe=gray!15]
\textbf{Answer to RQ2:} 
Both modules are essential for the performance of {\tool}. Adding examples retrieval improves 
{\tool} by 42.2\%, and introducing reasoning prompt  improves
{\tool} by 79.7\% in terms of SuccessRate@1.

\end{tcolorbox}

\subsection{Impacts of Different Examples (RQ3)}

\noindent \textbf{Experimental Design.} 
To answer this research question, we vary the number of examples from one to nine and compare three example selection methods as presented in Section~\ref{sec:exampleRetrieval}: BM-25~\cite{DBLP:journals/ftir/RobertsonZ09}, SBERT~\cite{DBLP:conf/emnlp/ReimersG19} and CodeT5~\cite{wang-etal-2021-codet5}. For BM-25, we implement with the gensim package~\cite{Gensim} by retrieving examples with the highest similarity from the training set. For dense retrieval methods, i.e., SBERT and CodeT5, we directly use these pre-trained models in the replication packages released by the authors without further tuning. Based on the text representations output by the pre-trained models, we select the examples presenting the highest cosine similarities in the training set.

\noindent \textbf{Results.} Fig.~\ref{fig:para_method} presents the results of {\tool} using various examples in method-level and class-level recommendation. We choose SuccessRate@3 and MAP@3 as the representative metrics to evaluate accuracy and ranking performance of {\tool} on APIBENCH-Q.

\noindent \textbf{Analyses.} 
We observe that the performance of {\tool} is largely affected by the selection of examples and the number of examples.
\uline{(1) SBERT proves to be an effective method for example retrieval.} 
Using SBERT for example selection, {\tool}'s performance demonstrates a substantial improvement. For example, in method-level API recommendation, SBERT achieves an average SuccessRate@3 of 0.52, marking an improvement of 15.12\% and 18.73\% compared to BM-25 and CodeT5, respectively. 
In class-level API recommendation, the average MAP@3 of SBERT is 0.59, showing improvements of 13.90\% and 14.07\% over BM-25 and CodeT5, respectively. One possible explanation is that SBERT is explicitly designed for understanding text semantics, enabling it to more accurately capture the semantic similarity between texts and thus making it more precise in identifying questions related to a query. In contrast, BM-25 is a traditional retrieval algorithm based on the bag-of-words model, often struggling to handle complex textual semantics. As for CodeT5, being a code pre-training model, it may have relatively weaker performance in natural language understanding. 
\uline{(2) The number of examples has varying effects on API recommendation.} Overall, a moderate increase in the number of examples can enhance performance. Taking method-level API recommendation as an example, when using SBERT, the performance of {\tool} continues to improve as the number of examples increases. However, in the case of BM25 and CodeT5, {\tool}'s performance peaks at four examples and then starts to decline. 
One possible explanation is that as the number of examples increases, the examples selected by SBERT have higher quality, while those selected by BM25 and CodeT5 may introduce more noise data. 
Additionally, we observe that the performance drops notably when there is only one example. This indicates that with only one example, the model lacks sufficient information for accurate recommendations. 

\begin{tcolorbox}[breakable,width=\linewidth-2pt,boxrule=0pt,top=1pt, bottom=1pt, left=4pt,right=4pt, colback=gray!15,colframe=gray!15]
\textbf{Answer to RQ3:} 
Selecting an appropriate number of examples and employing an effective example selection is crucial for the performance of
API recommendation.
\end{tcolorbox}

\subsection{Performance on Different Large Language Models (RQ4)}

\noindent \textbf{Experimental Design.} To answer this research question, we utilize two additional LLMs: ChatGPT (gpt-3.5-turbo)~\cite{chatgpt2022} and GPT-4 (gpt-4)~\cite{DBLP:journals/corr/abs-2303-08774}. We establish two settings: ``zero-shot LLM'' and ``APIGen-LLM''. In the zero-shot LLM setting, we do not provide any examples and only use the query as the input prompt for LLMs. In the APIGen-LLM setting, we use the output of our proposed \textit{example retrieval} and \textit{prompt construction} modules to construct the input prompt for LLMs.

\noindent \textbf{Results.} Fig.~\ref{fig:para_llm} presents the results of {\tool} with different LLMS at the method-level and class-level. Following RQ3, we use the same metrics to evaluate the performance of {\tool}, i.e., SuccessRate@3 and MAP@3.

\noindent \textbf{Analyses.} \uline{{\tool} can be effectively employed to improve the performance of various LLMs.} 
The experimental results indicate that {\tool} improves the performance of LLMs at both method-level and class-level, compared to LLMs with the zero-shot setting. 
For method-level recommendation, {\tool}-LLMs achieve substantial improvements of 49.87\% in average SuccessRate@3 and 60.29\% in average MAP@3. In class-level recommendation, {\tool} has a substantial positive effect on ChatGPT. The SuccessRate@3 increases from 0.323 to 0.609, showing an impressive 88.54\% improvement, and the MAP@3 increased from 0.275 to 0.551, showing a 100.36\% improvement. These results highlight the importance of using examples and reasoning prompts to enhance ChatGPT's performance in the API recommendation task. For GPT-4.0, relatively good results are achieved in both settings.

\begin{tcolorbox}[breakable,width=\linewidth-2pt,boxrule=0pt,top=1pt, bottom=1pt, left=4pt,right=4pt, colback=gray!15,colframe=gray!15]
\textbf{Answer to RQ4:} 
 {\tool} can enhance the performance of various LLMs, demonstrating its generalizability across different models.
\end{tcolorbox}

%% file: sections/5.0_main_table.tex
\begin{table*}[htbp]
\centering
\renewcommand\arraystretch{1.25}
\caption{Performance of {\tool} and the baseline approaches in method-level recommendation.}
\label{tab:comparison_method}
\begin{threeparttable}
\begin{tabular}{cc|ccc|ccc|c|ccc}
    \bottomrule
    \multicolumn{2}{c|}{\multirow{2}{*}{Method-level}} & \multicolumn{3}{c|}{SuccessRate@k} & \multicolumn{3}{c|}{MAP@k} & \multirow{2}{*}{MRR} & \multicolumn{3}{c}{NDCG@k} \\
    &  & Top-1 & Top-3 & Top-5 & Top-1 & Top-3 & Top-5 &  & Top-1 & Top-3 & Top-5 \\
    \hline
    \multicolumn{1}{c|}{\multirow{4}{*}{APIBENCH-Q}}  & DeepAPI  
    &0.01   &0.03   & 0.03  & 0.01  & 0.02   & 0.02   & 0.02   &  0.08   &  0.11   &  0.12    \\
    \multicolumn{1}{c|}{} & BIKER  
    &0.12   &0.23    & 0.29  & 0.12  & 0.16  & 0.18  &0.19  &0.27   & 0.32  & 0.35     \\
    \multicolumn{1}{c|}{} & CLEAR  
    &0.17   &0.31    &0.40   &0.17   &0.23  & 0.25  &0.25  &0.35   &0.46   &0.51      \\
    \cline{2-12}
    \multicolumn{1}{c|}{} & {\tool}  
    & \textbf{0.35* } & \textbf{0.50*}  & \textbf{0.55*} & \textbf{0.35*} & \textbf{0.42*}  & \textbf{0.43*}  & \textbf{0.43*}  & \textbf{0.54*}   & \textbf{0.59*}  & \textbf{0.60*}     \\

    \hline
    \multicolumn{1}{c|}{\multirow{4}{*}{BIKER-Dataset}} & DeepAPI    
    &0.01  &0.01   &0.01  &0.01  &0.01   &0.01 &0.01   &0.06    &0.09   &0.09      \\
    \multicolumn{1}{c|}{}  & BIKER  
    &0.43   & 0.66   & 0.75  & 0.35  & 0.47   & 0.50   &  0.55  &  0.71   & 0.73    &  0.74   \\
    \multicolumn{1}{c|}{}  & CLEAR  
    &0.48   & 0.65   & 0.71  &0.41   &0.51    & 0.52   &  0.57  & 0.72    & 0.75    &  0.76    \\
    \cline{2-12}
    \multicolumn{1}{c|}{} & {\tool}  
    & \textbf{0.53*}   & \textbf{0.70*}    & \textbf{0.77*}  & \textbf{0.43*}   & \textbf{0.54*}    & \textbf{0.56*}    & \textbf{0.62*}   & \textbf{0.73*}    &  \textbf{0.77*}   & \textbf{0.79*}     \\
    
    \toprule
    
    \end{tabular}
    \begin{tablenotes}
        \footnotesize
        \item * denotes statistically significant improvement (t-test with $p$-value $<$ 0.01) over the baseline approaches.
        \end{tablenotes}
    \end{threeparttable}
\end{table*}

\begin{table*}[htbp]
\centering
\renewcommand\arraystretch{1.25}
\caption{Performance of {\tool} and the baseline approaches in class-level recommendation.}
\label{tab:comparison_class}
    \begin{tabular}{cc|ccc|ccc|c|ccc}
    \bottomrule
    \multicolumn{2}{c|}{\multirow{2}{*}{Class-level}} & \multicolumn{3}{c|}{SuccessRate@k} & \multicolumn{3}{c|}{MAP@k} & \multirow{2}{*}{MRR} & \multicolumn{3}{c}{NDCG@k} \\
    &  & Top-1 & Top-3 & Top-5 & Top-1 & Top-3 & Top-5 &  & Top-1 & Top-3 & Top-5 \\
    \hline
    \multicolumn{1}{c|}{\multirow{5}{*}{APIBENCH-Q}} & RACK  
    &0.11   &0.20  & 0.23  & 0.11  & 0.15   & 0.16   & 0.16   & 0.11    & 0.17   & 0.18     \\
    \multicolumn{1}{c|}{} & DeepAPI  
    & 0.08  & 0.14   & 0.15   & 0.08  &0.10    & 0.11   & 0.11   & 0.08    & 0.11    & 0.12     \\
    \multicolumn{1}{c|}{} & BIKER  
    & 0.33  & 0.51   & 0.59  & 0.33  & 0.41  & 0.41   & 0.44   & 0.27    & 0.32    &  0.35    \\
    \multicolumn{1}{c|}{} & CLEAR  
    & 0.35  & 0.55   & 0.63  & 0.35  & 0.44   & 0.47   & 0.47   & 0.35    & 0.46    &  0.51    \\
    \cline{2-12}
    \multicolumn{1}{c|}{} & {\tool}  
    & \textbf{0.54*}  & \textbf{0.64*}  & \textbf{0.67*}  & \textbf{0.54*}  & \textbf{0.59*}   &  \textbf{0.59*}  &  \textbf{0.59*}  &  \textbf{0.54*}   & \textbf{0.59*}  &  \textbf{0.60*}    \\

    \hline
    
    \multicolumn{1}{c|}{\multirow{5}{*}{BIKER-Dataset}} & RACK  
    & 0.23  & 0.36   &0.38   &0.22   & 0.28   &  0.29  & 0.32   & 0.25    & 0.34    & 0.35     \\
    \multicolumn{1}{c|}{} & DeepAPI  
    &0.06   &0.10   & 0.12  & 0.06  & 0.07   & 0.08   & 0.09   &  0.06   &  0.09   &  0.09    \\
    \multicolumn{1}{c|}{}  & BIKER  
    &0.64   &0.83   &0.88   & 0.64   & 0.70   & 0.72   & 0.76   &  0.71   &  0.73  &  0.74   \\
    \multicolumn{1}{c|}{}  & CLEAR  
    &0.67   &0.80   & 0.85  & 0.65   & 0.68   & 0.71   & 0.73   & 0.72    & 0.75    & 0.76     \\
    \cline{2-12}
    \multicolumn{1}{c|}{} & {\tool}  
    & \textbf{0.73*}  & \textbf{0.87*}   & \textbf{0.95*}  & \textbf{0.66*}  &  \textbf{0.71*}  &  \textbf{0.73*}  & \textbf{0.79*}   &  \textbf{0.73*}   & \textbf{0.77*}    &  \textbf{0.79*}    \\
    
    \toprule
    
    \end{tabular}

\end{table*}

%% file: sections/5.1_aba_table.tex
\begin{table*}[htbp]
\centering
\renewcommand\arraystretch{1.25}
\caption{Performance of {\tool} and its variants in method-level recommendation and class-level recommendation.}
\label{tab:aba}
    \begin{tabular}{cc|ccc|ccc|ccc|ccc}
    \bottomrule
    \multicolumn{2}{c|}{\multirow{3}{*}{Ablation}}  & \multicolumn{6}{c|}{Method-level} & \multicolumn{6}{c}{Class-level} \\
     \cline{3-14}
     &  & \multicolumn{3}{c|}{SuccessRate@k} & \multicolumn{3}{c|}{MAP@k} & \multicolumn{3}{c|}{SuccessRate@k} & \multicolumn{3}{c}{MAP@k} \\
     &  & Top-1 & Top-3 & Top-5 & Top-1 & Top-3 & Top-5 & Top-1 & Top-3 & Top-5 & Top-1 & Top-3 & Top-5 \\
    \hline
    \multicolumn{1}{c|}{\multirow{3}{*}{APIBENCH-Q}} & w/o Example 
    & 0.24  & 0.38  & 0.44   & 0.24   & 0.30   & 0.32   & 0.36   &0.43     &  0.50   & 0.36  & 0.38    & 0.41     \\
    \multicolumn{1}{c|}{}  & w/o Reasoning  
    & 0.31  & 0.45  &  0.50  & 0.31   & 0.38   & 0.39   & 0.50   & 0.56    & 0.61    & 0.50   &  0.52   &  0.53    \\
    \cline{2-14}
    \multicolumn{1}{c|}{} & {\tool}  
    & \textbf{0.35}  & \textbf{0.50}  & \textbf{0.55}   & \textbf{0.35}   & \textbf{0.42}   & \textbf{0.43}   & \textbf{0.54}   & \textbf{0.64}    & \textbf{0.67}  &  \textbf{0.54}  &  \textbf{0.59}   &  \textbf{0.59}    \\
    \hline
    \multicolumn{1}{c|}{\multirow{3}{*}{BIKER-Dataset}} & w/o Example  
    & 0.37  & 0.54  & 0.61   & 0.29   & 0.39   & 0.44   & 0.58   & 0.72    &  0.77   &  0.52  &   0.60  &  0.61   \\
    \multicolumn{1}{c|}{}  & w/o Reasoning  
    & 0.49  & 0.67  & 0.73    & 0.40  & 0.49   &0.52    & 0.69   &  0.75   & 0.86    &0.60    & 0.68    & 0.70      \\
    \cline{2-14}
    \multicolumn{1}{c|}{} & {\tool}  
    & \textbf{0.54} & \textbf{0.70}  & \textbf{0.77}   & \textbf{0.43}   & \textbf{0.54}   & \textbf{0.56}   & \textbf{0.73}   & \textbf{0.87}    & \textbf{0.95}  & \textbf{0.66} & \textbf{0.71}  & \textbf{0.73}     \\
    \toprule
    \end{tabular}
\end{table*}

%% file: sections/5.2_para_figure.tex
\definecolor{c1}{RGB}{255,196,115} 
\definecolor{c2}{RGB}{178,37,42} 
\definecolor{c3}{RGB}{103,150,118} 
\definecolor{c4}{RGB}{181,181,181} 


\pgfplotstableread[row sep=\\,col sep=&]{
	k & BM-25 & SBERT & CodeT5   \\
	1 & 0.447 & 0.436 & 0.442  \\
	2 & 0.451 & 0.476 & 0.447 \\
	3 & 0.453 & 0.500 & 0.464 \\
	4 & 0.449 & 0.533 & 0.469 \\
 	5 & 0.453 & 0.529 & 0.440 \\
        6 & 0.473 & 0.536 & 0.451 \\
        7 & 0.473 & 0.578 & 0.427 \\
        8 & 0.462 & 0.573 & 0.418 \\
        9 & 0.440 & 0.560 & 0.418 \\
}\MASR

\pgfplotstableread[row sep=\\,col sep=&]{
	k & BM-25 & SBERT & CodeT5   \\
	1 & 0.371 & 0.353 & 0.364  \\
	2 & 0.362 & 0.397 & 0.355 \\
	3 & 0.373 & 0.425 & 0.374 \\
	4 & 0.367 & 0.436 & 0.370 \\
 	5 & 0.380 & 0.445 & 0.351 \\
        6 & 0.377 & 0.446 & 0.364 \\
        7 & 0.384 & 0.487 & 0.350 \\
        8 & 0.382 & 0.476 & 0.342 \\
        9 & 0.368 & 0.500 & 0.338 \\
}\MAMA

\pgfplotstableread[row sep=\\,col sep=&]{
	k & BM-25 & SBERT & CodeT5   \\
	1 & 0.520  & 0.551 & 0.520  \\
	2 & 0.542 & 0.591 & 0.567 \\
	3 & 0.560 & 0.604 & 0.573 \\
	4 & 0.564 & 0.649 & 0.578 \\
 	5 & 0.569 & 0.649 & 0.571 \\
        6 & 0.596 & 0.667 & 0.587 \\
        7 & 0.580 & 0.707 & 0.578 \\
        8 & 0.576 & 0.698 & 0.558 \\
        9 & 0.576 & 0.716 & 0.580 \\
}\CASR

\pgfplotstableread[row sep=\\,col sep=&]{
	k & BM-25 & SBERT & CodeT5   \\
	1 & 0.477 & 0.501 & 0.474  \\
	2 & 0.492 & 0.541 & 0.520 \\
	3 & 0.513 & 0.561 & 0.522 \\
	4 & 0.517 & 0.585 & 0.521 \\
 	5 & 0.530 & 0.587 & 0.514 \\
        6 & 0.535 & 0.602 & 0.532 \\
        7 & 0.526 & 0.633 & 0.525 \\
        8 & 0.528 & 0.640 & 0.509 \\
        9 & 0.530 & 0.644 & 0.524 \\
}\CAMA

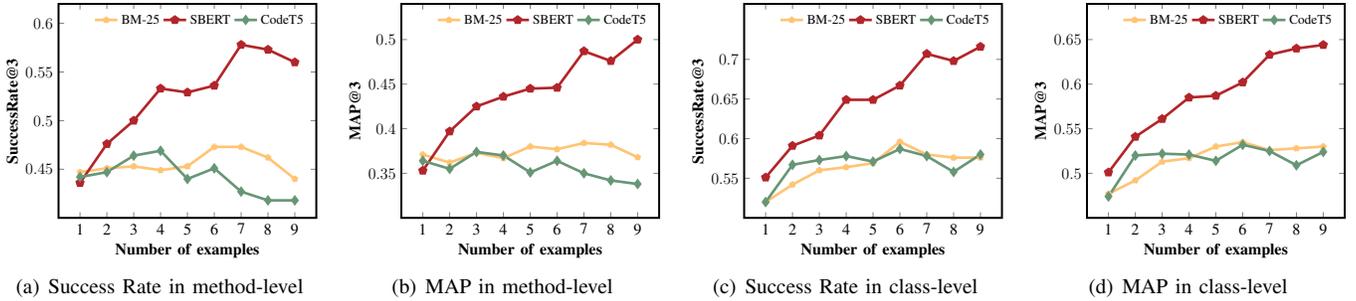
\begin{figure*}[h]
	\centering		
	\subfigure[Success Rate in method-level]{
		\begin{tikzpicture}[scale=0.50]
			\begin{axis}[
			    legend style = {
				    legend columns=3,
				    draw=none,
				},
				xtick = {1,2,3,4,5,6,7,8,9},
				xticklabels = {1,2,3,4,5,6,7,8,9},
				ymin=0.40,ymax=0.62,
				ytick = {0.45, 0.50, 0.55, 0.60},
				mark size=2.5pt, 
				ylabel={\large \bf SuccessRate@3},
				xlabel={\large \bf Number of examples}, 
				ticklabel style={font=\large},
				every axis plot/.append style={line width = 2pt},
				every axis/.append style={line width = 1.5pt},
				]
				\addplot [mark=star,color=c1] table[x=k,y=BM-25]{\MASR};
				\addplot [mark=pentagon*,color=c2] table[x=k,y=SBERT]{\MASR};
				\addplot [mark=diamond*,color=c3] table[x=k,y=CodeT5]{\MASR};
				
				\legend{BM-25, SBERT, CodeT5}
			\end{axis}
		\end{tikzpicture}
		\label{fig:res/pre.SHS}
	}
	\subfigure[MAP in method-level]{
		\begin{tikzpicture}[scale=0.50]
			\begin{axis}[
			    legend style = {
				    legend columns=3,
				    draw=none,
				},
				xtick = {1,2,3,4,5,6,7,8,9},
				xticklabels = {1,2,3,4,5,6,7,8,9},
				ymin=0.30,ymax=0.54,
				ytick = {0.35, 0.4, 0.45, 0.50},
				mark size=2.5pt, 
				ylabel={\large \bf MAP@3},
				xlabel={\large \bf Number of examples}, 
				ticklabel style={font=\large},
				every axis plot/.append style={line width = 2pt},
				every axis/.append style={line width = 1.5pt},
				]
				\addplot [mark=star,color=c1] table[x=k,y=BM-25]{\MAMA};
				\addplot [mark=pentagon*,color=c2] table[x=k,y=SBERT]{\MAMA};
				\addplot [mark=diamond*,color=c3] table[x=k,y=CodeT5]{\MAMA};
				
				\legend{BM-25, SBERT, CodeT5}
			\end{axis}
		\end{tikzpicture}
		\label{fig:res/pre.SHS}
	}
	\subfigure[Success Rate in class-level]{
		\begin{tikzpicture}[scale=0.50]
			\begin{axis}[
			    legend style = {
				    legend columns=3,
				    draw=none,
				},
				xtick = {1,2,3,4,5,6,7,8,9},
				xticklabels = {1,2,3,4,5,6,7,8,9},
				ymin=0.50,ymax=0.77,
				ytick = {0.55, 0.60, 0.65, 0.70},
				mark size=2.5pt, 
				ylabel={\large \bf SuccessRate@3},
				xlabel={\large \bf Number of examples}, 
				ticklabel style={font=\large},
				every axis plot/.append style={line width = 2pt},
				every axis/.append style={line width = 1.5pt},
				]

				\addplot [mark=star,color=c1] table[x=k,y=BM-25]{\CASR};
				\addplot [mark=pentagon*,color=c2] table[x=k,y=SBERT]{\CASR};
				\addplot [mark=diamond*,color=c3] table[x=k,y=CodeT5]{\CASR};
				
				\legend{BM-25, SBERT, CodeT5}
			\end{axis}
		\end{tikzpicture}
		\label{fig:res/pre.SHS}
	}
        \subfigure[MAP in class-level]{
		\begin{tikzpicture}[scale=0.50]
			\begin{axis}[
			    legend style = {
				    legend columns=3,
				    draw=none,
				},
				xtick = {1,2,3,4,5,6,7,8,9},
				xticklabels = {1,2,3,4,5,6,7,8,9},
				ymin=0.45,ymax=0.69,
				ytick = {0.50, 0.55, 0.60, 0.65},
				mark size=2.5pt, 
				ylabel={\large \bf MAP@3},
				xlabel={\large \bf Number of examples}, 
				ticklabel style={font=\large},
				every axis plot/.append style={line width = 2pt},
				every axis/.append style={line width = 1.5pt},
				]
				\addplot [mark=star,color=c1] table[x=k,y=BM-25]{\CAMA};
				\addplot [mark=pentagon*,color=c2] table[x=k,y=SBERT]{\CAMA};
				\addplot [mark=diamond*,color=c3] table[x=k,y=CodeT5]{\CAMA};
				
				\legend{BM-25, SBERT, CodeT5}
			\end{axis}
		\end{tikzpicture}
		\label{fig:res/pre.SHS}
	}
	\caption{Experimental results with different examples in method-level recommendation and class-level recommendation.}
	\label{fig:para_method}
\end{figure*}

\pgfplotstableread[row sep=\\,col sep=&]{
      datasets  & without & with   \\
    2 & 0.38  & 0.50    \\ 
    4 & 0.256 & 0.516   \\
    6 & 0.431 & 0.502  \\
}\MSR

\pgfplotstableread[row sep=\\,col sep=&]{
      datasets & without & with    \\
    2 & 0.30 & 0.42    \\
    4 & 0.203 & 0.45   \\
    6 & 0.349 & 0.416  \\
}\MMA

\pgfplotstableread[row sep=\\,col sep=&]{
      datasets  & without & with   \\
    2 & 0.43 & 0.64    \\
    4 & 0.323 & 0.609   \\
    6 & 0.549 & 0.642  \\
}\CSR	

\pgfplotstableread[row sep=\\,col sep=&]{
      datasets & without & with    \\
    2 & 0.38  & 0.59     \\
    4 & 0.275 & 0.551   \\
    6 & 0.474 & 0.565  \\
}\CMA

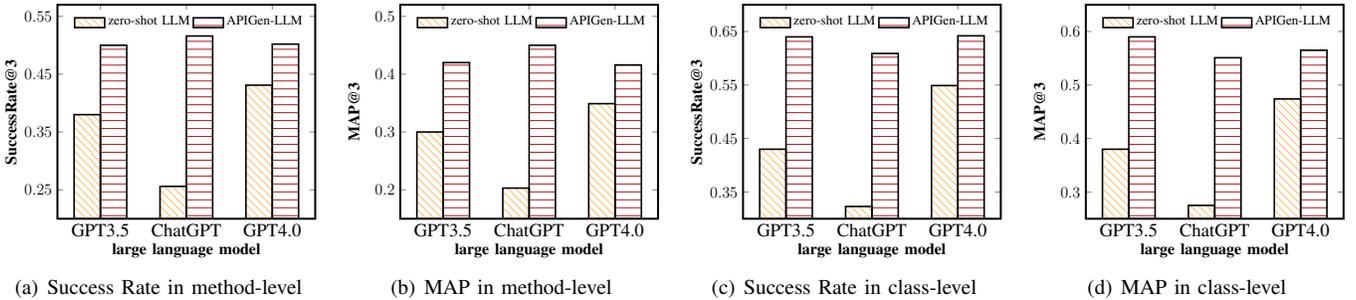
\begin{figure*}[h]	

	\subfigure[Success Rate in method-level]{
    	\begin{tikzpicture}[scale=0.50]
    		\begin{axis}[
    	    	ybar=0pt,
    			bar width=0.7cm,
    			xlabel={\large \bf large language model}, 
                    xmin=1.0,xmax=7.0,
                    xtick=data,	xticklabels={\Large GPT3.5,\Large ChatGPT, \Large GPT4.0},
                legend style = {
				    legend columns=-1,
				    draw=none,
				},
				legend image code/.code={
                    \draw [#1] (0cm,-0.18cm) rectangle (0.8cm,0.08cm); },
    			ytick = {0.25, 0.35, 0.45, 0.55},
    			ymin=0.2,ymax=0.57,
    			tick align=inside,
    			ticklabel style={font=\large},
   			  every axis plot/.append style={line width = 1.2pt},
    			every axis/.append style={line width = 1.5pt},
    			ylabel={\large \textbf{SuccessRate@3}},
    			]
    			\addplot[pattern=north west lines, pattern color=c1] table[x=datasets,y=without]{\MSR};
    			\addplot[pattern = horizontal lines ,pattern color=c2] table[x=datasets,y=with]{\MSR};
    			\legend{zero-shot LLM, {\tool}-LLM}
    
    		\end{axis}
    	\end{tikzpicture}
    	\label{fig:res-para-shs}
	}
	\subfigure[MAP in method-level]{
    	\begin{tikzpicture}[scale=0.50]
    		\begin{axis}[
    	    	ybar=0pt,
    			bar width=0.7cm,
    			xlabel={\large \bf large language model}, 
                    xmin=1.0,xmax=7.0,
                    xtick=data,	xticklabels={\Large GPT3.5,\Large ChatGPT, \Large GPT4.0},
                legend style = {
				    legend columns=-1,
				    draw=none,
				},
				legend image code/.code={
                    \draw [#1] (0cm,-0.18cm) rectangle (0.8cm,0.08cm); },
    			ytick = {0.20, 0.30, 0.40, 0.50},
    			ymin=0.15,ymax=0.52,
    			tick align=inside,
    			ticklabel style={font=\large},
   			  every axis plot/.append style={line width = 1.2pt},
    			every axis/.append style={line width = 1.5pt},
    			ylabel={\large \textbf{MAP@3}},
    			]
    			\addplot[pattern=north west lines, pattern color=c1] table[x=datasets,y=without]{\MMA};
    			\addplot[pattern = horizontal lines ,pattern color=c2] table[x=datasets,y=with]{\MMA};
    			\legend{zero-shot LLM, {\tool}-LLM}
    
    		\end{axis}
    	\end{tikzpicture}
    	\label{fig:res-para-shs}
	}
	\subfigure[Success Rate in class-level]{
    	\begin{tikzpicture}[scale=0.50]
    		\begin{axis}[
    	    	ybar=0pt,
    			bar width=0.7cm,
    			xlabel={\large \bf large language model}, 
                    xmin=1.0,xmax=7.0,
                    xtick=data,	xticklabels={\Large GPT3.5,\Large ChatGPT, \Large GPT4.0},
                legend style = {
				    legend columns=-1,
				    draw=none,
				},
				legend image code/.code={
                    \draw [#1] (0cm,-0.18cm) rectangle (0.8cm,0.08cm); },
    			ytick = {0.35, 0.45, 0.55, 0.65},
    			ymin=0.3,ymax=0.7,
    			tick align=inside,
    			ticklabel style={font=\large},
   			  every axis plot/.append style={line width = 1.2pt},
    			every axis/.append style={line width = 1.5pt},
    			ylabel={\large \textbf{SuccessRate@3}},
    			]
    			\addplot[pattern=north west lines, pattern color=c1] table[x=datasets,y=without]{\CSR};
    			\addplot[pattern = horizontal lines ,pattern color=c2] table[x=datasets,y=with]{\CSR};
    			\legend{zero-shot LLM, {\tool}-LLM}
    
    		\end{axis}
    	\end{tikzpicture}
    	\label{fig:res-para-shs}
	}
        \subfigure[MAP in class-level]{
    	\begin{tikzpicture}[scale=0.50]
    		\begin{axis}[
    	    	ybar=0pt,
    			bar width=0.7cm,
    			xlabel={\large \bf large language model}, 
                    xmin=1.0,xmax=7.0,
                    xtick=data,	xticklabels={\Large GPT3.5,\Large ChatGPT, \Large GPT4.0},
                legend style = {
				    legend columns=-1,
				    draw=none,
				},
				legend image code/.code={
                    \draw [#1] (0cm,-0.18cm) rectangle (0.8cm,0.08cm); },
    			ytick = {0.30, 0.40, 0.50, 0.60},
    			ymin=0.25,ymax=0.65,
    			tick align=inside,
    			ticklabel style={font=\large},
   			  every axis plot/.append style={line width = 1.2pt},
    			every axis/.append style={line width = 1.5pt},
    			ylabel={\large \textbf{MAP@3}},
    			]
    			\addplot[pattern=north west lines, pattern color=c1] table[x=datasets,y=without]{\CMA};
    			\addplot[pattern = horizontal lines ,pattern color=c2] table[x=datasets,y=with]{\CMA};
    			\legend{zero-shot LLM, {\tool}-LLM}
    
    		\end{axis}
    	\end{tikzpicture}
    	\label{fig:res-para-shs}
	}
	
        \caption{Experimental results with different large language models in method-level recommendation and class-level recommendation.}
        \label{fig:para_llm}

\end{figure*}

%% file: sections/6_discussion.tex
\subsection{Case Study}

\begin{figure}
    \centering
    \includegraphics[scale=0.42]{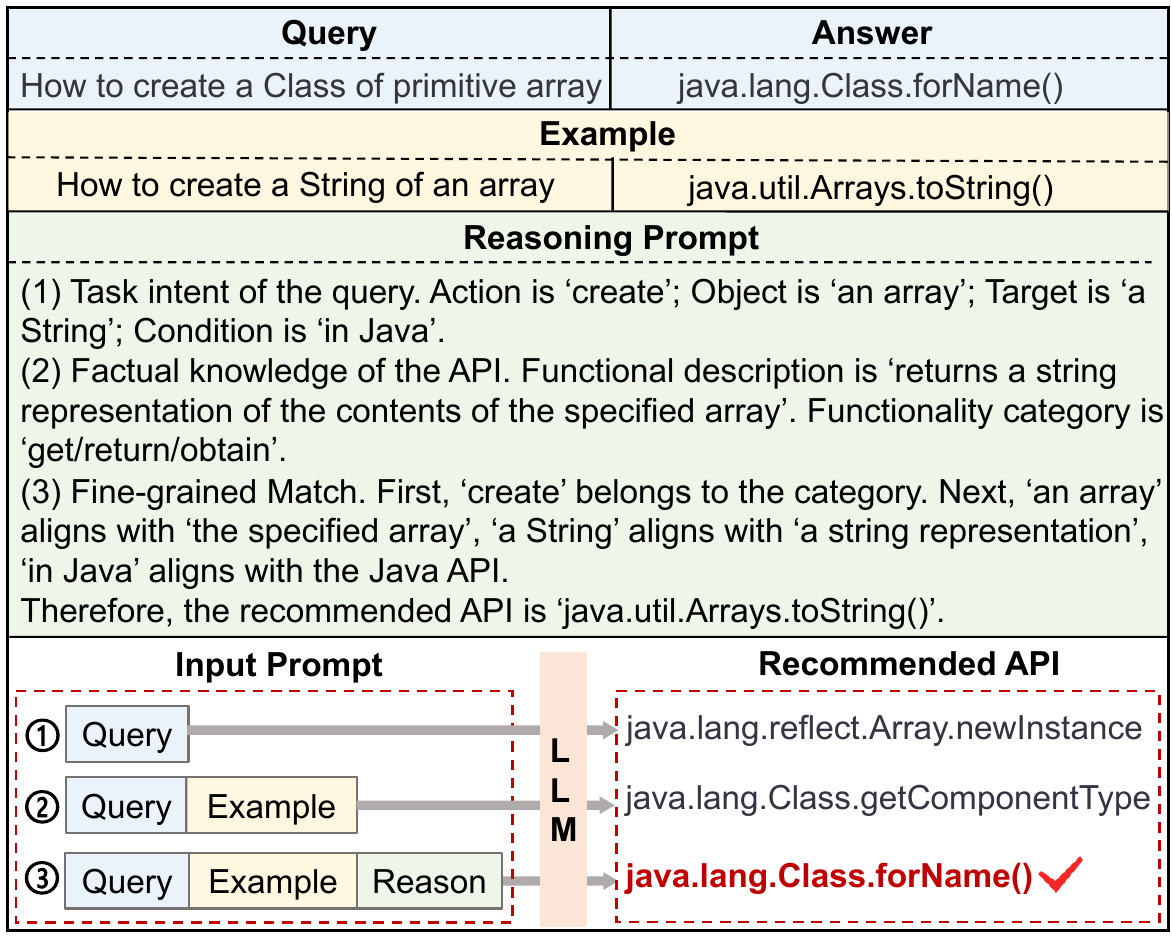}
    \caption{A case study of {\tool} with three different input prompts.}
    \label{fig:case}
\end{figure}

To explore how the example and reasoning prompt affect recommendation results of {\tool}, we conduct a case study by using the query ``\textit{How to create a Class of primitive array}'' shown in Fig.~\ref{fig:icl}, which is a common programming problem about creating a class with primitive data type arrays. 
As shown in Fig.~\ref{fig:case}, we use the SBERT method to select the top-1 similar post for the query from training data as an example, i.e., ``\textit{$<$How to create a String of an array, java.util.Arrays.toString()$>$}''. Next, we generate the reasoning prompt for the selected post following the steps in Section~\ref{sec:prompt}. Finally, we design three input prompts for LLMs in {\tool} by leveraging the query, the example, and the reasoning prompt. 
By observing the generated results, we analyze the impact of the example and reasoning prompt on API recommendation. 

\textbf{The impact of example.} For the first input prompt with only a given query, the LLM recommends the API: ``\textit{java.lang.reflect.Array.newInstance()}'', which is used to create a new array instance, not to create a new class. This error indicates that the LLM just focuses on the keyword ``Array'' without understanding the overall semantics of the query.
For the second input prompt, by adding an example, the recommendation API is: ``\textit{java.lang.Class.getComponentType()}'', which is used to retrieve the component type of an array, rather than creating a new class.    
Compared to the first recommended API, the second one is closer to the answer since it predicts the correct class.
This indicates the LLM learns how to understand the query from the example, thus paying more attention to the ``class'' keyword within the query. However, due to a lack of sufficient knowledge about Java classes and methods, it still generates an incorrect API.

\textbf{The impact of reasoning prompt.} For the third input prompt, by adding an example and a reasoning prompt, the generated result is ``\textit{java.lang.Class.forName()}'', which is correct. This API method is used to dynamically load and return a reference to a class at runtime. Compared to the second recommended API, {\tool} not only generates the correct class but also provides the correct method. This indicates that the reasoning prompt helps APIGen match the intent of the query and the knowledge of API, thus providing accurate API recommendation. 

This case study shows the importance of examples and reasons in API recommendations. The example provides learning context, while the reasoning prompt offers guidance to recommend APIs. By introducing both example and reason prompt, {\tool} comprehensively understands the query and makes predicted APIs better meet the programming requirement.

\subsection{Threats to Validity}

We identify four threats to the validity of our study:

\textbf{Potential data leakage.}
In our study, we conduct experiments using the API of OpenAI's GPT-3.5, which is a closed-source model, and its parameters and training data are not publicly available, raising concerns regarding potential data leaks. 
However, our experiments clearly show that {\tool} performs poorly without examples, indicating a low probability of direct memorization. Thus, we believe that the risk of data leakage in our study is minimal.

\textbf{The generalizability of our experimental results.}
In our work, we evaluate the performance of {\tool} on Java datasets. It can also be adapted to other programming languages like Python, since our approach relies solely on the natural language, i.e., query intent and API description.

\textbf{The design of prompt template.} 
In this work, we present only a single template illustrating how to perform reasoning based on the task intent of a query and factual knowledge of an API. While this demonstrates the effectiveness of reasoning prompts, we will develop more templates to further investigate the performance of {\tool} in the future.

\textbf{The selection of models.} In this paper, we select three LLMs for experiments. However, it's worth noting that there are other LLMs available, including Incoder~\cite{DBLP:conf/iclr/FriedAL0WSZYZL23} and CodeGen~\cite{DBLP:conf/iclr/NijkampPHTWZSX23}. In the future, we plan to conduct experiments with a more diverse range of LLMs to explore the applicability of our framework more comprehensively.

%% file: sections/7_relatedwork.tex
\subsection{API Recommendation}
The existing works on API recommendation include two categories: retrieval-based  and learning-based methods. 

\textbf{Retrieval-based methods.}
McMillan \etal propose portfolio~\cite{DBLP:conf/icse/McMillanGPXF11}, an API recommendation tool that supports programmers in finding relevant code snippets that implement high-level requirements reflected in query.
%
%
Rahman \etal propose RACK~\cite{DBLP:conf/wcre/RahmanRL16} to extract keyword-API relation and find relevant API classes based on the crowdsourced knowledge collected on Stack Overflow.
Huang \etal propose BIKER~\cite{DBLP:conf/kbse/HuangXXLW18} to obtain API candidates by calculating the similarity between queries and official documentations as well the Stack Overflow posts.
Wei \etal propose CLEAR~\cite{DBLP:conf/icse/WeiHH0022} to select a set of candidate Stack Overflow posts based on BERT sentence embedding similarity and re-ranks them using a BERT-based classification model to recommend the top-N APIs.
Different with these retrieval-based methods, {\tool} has a powerful representation capability, making it better understand queries and API documentation, which is benefited from the extensive text encoded in LLMs.

\textbf{Learning-based methods.}
Gu \etal propose DeepAPI~\cite{DBLP:conf/sigsoft/GuZZK16}, which is the first approach to introduce a deep learning model to recommend API sequences. It reformulates API recommendation task as a query-API translation problem and uses an RNN Encoder-Decoder model.
Ling \etal propose GeAPI~\cite{DBLP:journals/jcst/LingZLX19} to automatically construct API graphs based on source code and leverages graph embedding techniques for API representation. Given a query, it searches relevant subgraphs on the original graph and recommends them to developers. 
Zhou \etal propose BRAID~\cite{DBLP:journals/tse/ZhouYCHMG22} to boost API recommendation performance by leveraging learning-to-rank and active learning techniques. 
Compared to these learning-based methods, {\tool} can make effective recommendations with only a few examples via ICL, without requiring a large amount of labeled data.

\subsection{API Usage Pattern Mining}

The works on API usage pattern mining usually utilize traditional statistical methods to capture usage patterns from API co-occurrence or leverage deep learning models to automatically learn the potential usage patterns from a large code corpus. 
Zhong \etal propose MAPO~\cite{DBLP:conf/ecoop/ZhongXZPM09} to cluster and mine API usage patterns from open source repositories, and then recommends the relevant usage patterns to developers. 
Wang \etal improve MAPO and build UP-Miner~\cite{DBLP:conf/msr/WangDZCXZ13} by utilizing a new algorithm based on SeqSim to cluster the API sequences.
Nguyen \etal propose APIREC~\cite{DBLP:conf/sigsoft/NguyenHCNMRND16}, which uses fine-grained code changes and the corresponding changing contexts to recommend APIs.
Fowkes \etal propose PAM~\cite{DBLP:conf/sigsoft/FowkesS16} to mine API usage patterns through an almost parameter-free probabilistic algorithm and uses them to recommend APIs. 
Nguyen \etal propose a graph-based language model GraLan~\cite{DBLP:conf/icse/NguyenN15} to recommend API usages.

%% file: sections/8_conclusion.tex
In this paper, we introduce {\tool}, a generative API recommendation approach through enhanced in-context learning. {\tool} incorporates fine-grained matching between the query's task intent and APIs' factual knowledge into large language models via a novel prompt design. This enables the large language models to better understand queries and generate more suitable API recommendation. Experimental results demonstrate that {\tool} outperforms state-of-the-art
methods in both method-level and class-level API recommendation.